\documentclass[10pt,conference]{IEEEtran}
\IEEEoverridecommandlockouts

\usepackage{cite}
\usepackage{amsmath,amssymb,amsfonts}
\usepackage{algorithmic}
\usepackage{graphicx}
\usepackage{textcomp}
\usepackage{xcolor}
\usepackage{url}

\usepackage{hyperref}

\usepackage{booktabs}

\usepackage{listings} % For code-style formatting

\newcommand{\TheName}{\textsc{AutoPatch}}

\def\BibTeX{{\rm B\kern-.05em{\sc i\kern-.025em b}\kern-.08em
    T\kern-.1667em\lower.7ex\hbox{E}\kern-.125emX}}
\begin{document}

% \title{Enabling Efficient Code Optimization through Context-Aware
%   Retrieval-Augmented Generation\\

\title{Optimizing Code Runtime Performance through Context-Aware Retrieval-Augmented Generation}
% {\footnotesize \textsuperscript{*}Note: Sub-titles are not captured for https://ieeexplore.ieee.org  and

\author{
\IEEEauthorblockN{
Manish Acharya\textsuperscript{†*}\thanks{*Equal contribution; authors listed in alphabetical order.},
Yifan Zhang\textsuperscript{†‡*},
Kevin Leach\textsuperscript{†‡},
Yu Huang\textsuperscript{†‡}
}
\IEEEauthorblockA{\textsuperscript{†}Department of Computer Science, Vanderbilt University, Nashville, TN, USA\\}
\IEEEauthorblockA{\textsuperscript{‡}Institute for Software Integrated Systems, Vanderbilt University, Nashville, TN, USA\\}
}

% \author{
% \IEEEauthorblockN{
% Manish Acharya\textsuperscript{†*}\thanks{*Equal contribution; authors listed in alphabetical order.},
% Yifan Zhang\textsuperscript{†*},
% Kevin Leach\textsuperscript{†},
% Yu Huang\textsuperscript{†}
% }
% \IEEEauthorblockA{\textsuperscript{†}Department of Computer Science, Vanderbilt University, Nashville, TN, USA\\}
% }

% \author{Anonymous Author(s)}

% \author{\IEEEauthorblockN{1\textsuperscript{st} Manish Acharya}
% \IEEEauthorblockA{\textit{Department of Computer Science} \\
% \textit{Vanderbilt University}\\
% Nashville, Tennessee\\
% manish.acharya@vanderbilt.edu}
% \and
% \IEEEauthorblockN{2\textsuperscript{nd} Yifan Zhang}
% \IEEEauthorblockA{\textit{Department of Computer Science} \\
% \textit{Vanderbilt University}\\
% Nashville, Tennessee\\
% yifan.zhang.2@vanderbilt.edu}
% \and
% \IEEEauthorblockN{3\textsuperscript{rd} Kevin Leach}
% \IEEEauthorblockA{\textit{Department of Computer Science} \\
% \textit{Vanderbilt University}\\
% Nashville, Tennessee\\
% kevin.leach@vanderbilt.edu}
% \and
% \IEEEauthorblockN{4\textsuperscript{th} Yu Huang}
% \IEEEauthorblockA{\textit{Department of Computer Science} \\
% \textit{Vanderbilt University}\\
% Nashville, Tennessee\\
% yu.huang@vanderbilt.edu}
% }

\maketitle

\begin{abstract}
% Optimizing software performance through automated code refinement offers an opportunity to enhance execution speed and efficiency. This study presents a Retrieval-Augmented Generation (RAG) approach for large language models (LLMs) aimed at improving program performance by generating optimized C++ code. By leveraging Control Flow Graph (CFG) analysis and differences between the CFGs of original and optimized code, our method identifies bottlenecks, eliminates redundancies, and streamlines execution paths. CFG-based insights and relevant optimization examples are retrieved as context, enabling the LLM to generate improved code aligned with structural and algorithmic refinements. While our experiments demonstrate improvements over baseline methods in lexical similarity metrics, future work will include runtime validation to confirm the correctness and effectiveness of the optimizations. This approach highlights the potential of RAG-enhanced LLMs in program optimization tasks.

Optimizing software performance through automated code refinement offers a promising avenue for enhancing execution speed and efficiency. Despite recent advancements in LLMs, a significant gap remains in their ability to perform in-depth program analysis. This study introduces \TheName{}, an in-context learning approach designed to bridge this gap by enabling LLMs to automatically generate optimized code. Inspired by how programmers learn and apply knowledge to optimize software, \TheName{} incorporates three key components: (1) an analogy-driven framework to align LLM optimization with human cognitive processes, (2) a unified approach that integrates historical code examples and CFG analysis for context-aware learning, and (3) an automated pipeline for generating optimized code through in-context prompting. Experimental results demonstrate that \TheName{} achieves a 7.3\% improvement in execution efficiency over GPT-4o across common generated executable code, highlighting its potential to advance automated program runtime optimization.

% By leveraging Control Flow Graph (CFG) analysis and differences between the CFGs of original and optimized code, our method identifies bottlenecks, eliminates redundancies, and streamlines execution paths. CFG-based insights and relevant optimization examples are retrieved as context, enabling the LLM to generate improved code aligned with structural and algorithmic refinements. While our experiments demonstrate improvements over baseline methods in lexical similarity metrics, future work will include runtime validation to confirm the correctness and effectiveness of the optimizations. This approach highlights the potential of RAG-enhanced LLMs in program optimization tasks.

\end{abstract}


\section{Introduction}

Optimizing program performance is increasingly vital as hardware advancements plateau and computational demands rise~\cite{li2024utilizing}. While traditional compilers excel at tasks like instruction scheduling and register allocation, higher-level optimizations such as restructuring logic, refining control flows, and addressing inefficiencies still depend heavily on human expertise~\cite{shypula2023learning}. Recent progress in large language models (LLMs) has introduced new opportunities for tackling these challenges~\cite{gao2024search,liu2023jarvix}, leveraging their ability to understand and generate complex code. At the same time, fully harnessing contextual and structural insights for LLM-driven optimization remains challenging, as current methods often fall short of emulating the nuanced analysis and dynamic application of improvements performed by human programmers~\cite{qiu2024efficient}.

To bridge this gap, LLMs must address key challenges, including identifying structural inefficiencies, adapting to diverse code contexts, and dynamically refining execution paths~\cite{chew2023llm}. Traditional methods rely heavily on manual analysis and predefined heuristics, limiting scalability and adaptability~\cite{fan2023static}. LLMs, however, offer the potential to overcome these limitations by learning from historical examples and applying optimization patterns in real-time~\cite{wadhwa2023frustrated,nam2024using}. While promising, the potential for LLMs to emulate human-like reasoning by leveraging program structure and contextual retrieval remains largely untapped. Exploring this avenue could enable more adaptive and scalable optimization techniques, seamlessly combining structural and semantic insights to address complex programming challenges.

Building on these opportunities, we highlight the importance of studying how human programmers optimize code, focusing on contextual understanding and leveraging historical knowledge. To address this, we propose \TheName{}, a context-aware framework enabling LLMs to perform in-depth program optimization. \TheName{} includes three components: (1) Formalizing methods to incorporate domain knowledge by studying programmer behavior and identifying inefficiencies. (2) Integrating context analysis with augmented retrieval to enable LLMs to learn from past optimizations. (3) Using enriched context, the LLM generates optimized code by combining learned patterns with historical insights. Evaluated on the IBM Project CodeNet dataset~\cite{puri2021codenet}, \TheName{} achieves a 7.3\% improvement in runtime performance over GPT-4o, bridging manual expertise and automated optimization.

\TheName{} highlights the potential for a unified LLM-powered pipeline for program analysis and optimization. Future advancements can expand its scope by incorporating complex retrieval mechanisms and neural program analysis, addressing more sophisticated programming challenges.

\begin{figure*}[htbp]
    \centering
    \includegraphics[width=\textwidth]{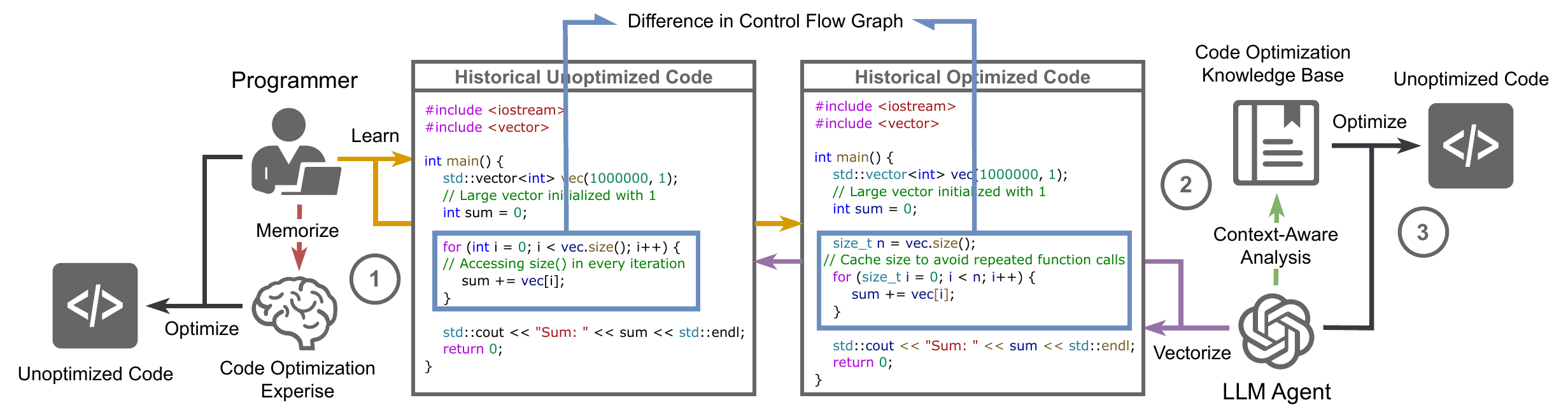}
    \caption{Overview of the \TheName{} Workflow: \TheName{} mimics human cognitive processes to optimize code using historical examples and CFG analysis. \textcircled{1} represents the programmer learning phase, extracting insights from historical code and CFG differences. \textcircled{2} integrates CFG analysis and optimization patterns into a context-aware LLM. \textcircled{3} applies this knowledge to generate optimized code for new inputs.}
    \label{fig:autopatch}
\end{figure*}

\section{Methodology}

This section presents the methodology for optimizing code using a context-aware framework inspired by expertise-driven analysis, as shown in Fig.~\ref{fig:autopatch}. The framework consists of three steps: \textcircled{1} formalizing optimization strategies based on human expertise to identify inefficiencies; \textcircled{2} leveraging LLMs to implement a retrieval pipeline that dynamically integrates historical examples; and \textcircled{3} designing context-enriched prompts combining human strategies with LLM-driven insights to produce optimized code.

\subsection{Studying Human Behavior in Code Optimization}

Human programmers optimize code by analyzing control flow structures and identifying inefficiencies. This process involves comparing the unoptimized and optimized versions of code to extract actionable insights, providing structured guidance for refinement.

\subsubsection{Understanding Control Flow through Programmer Analysis}  
Programmers start by examining the logical flow and decision points within code, focusing on sequences of operations and transitions to identify inefficiencies. This intuitive mapping of execution paths uncovers redundant loops, unnecessary branches, or suboptimal operations. Translating this process into CFG formalizes their analysis, enabling a structured approach to pinpoint bottlenecks and refine execution paths systematically.

\subsubsection{Extracting Optimization Patterns from Insights}  
After identifying inefficiencies, programmers compare unoptimized and optimized code to uncover transformations like simplified logic, refined control flows, and optimized loops. Recurring patterns, such as loop unrolling and redundancy elimination, are distilled into reusable strategies, forming a knowledge base that aligns with Retrieval-Augmented Generation (RAG) to guide future optimizations.

\subsubsection{Generalizing Knowledge for Actionable Strategies}  
Beyond immediate fixes, programmers abstract insights into systematic strategies, ensuring their applicability to diverse scenarios. This iterative process of analysis, refinement, and abstraction creates a feedback loop that formalizes human expertise into structured methodologies.

\subsection{Leveraging Historical Insights for LLM Optimization}

Building on human expertise in code optimization, we develop a framework combining CFG Diff Analysis and RAG to guide LLMs in optimizing target code (\( C_t \)). Preprocessed examples (\( E_r \)) are stored in a vector database for efficient retrieval and context-aware refinement.

\subsubsection{CFG Diff as Analytical Basis}  
Given the CFGs of the original (\( G_o \)) and optimized (\( G_p \)) code, we compute the difference, \( \Delta G = G_p - G_o \), capturing:
\[
\Delta G = (\Delta S, \Delta F, \Delta C),
\]
where \( \Delta S \) represents structural changes (e.g., added or removed blocks), \( \Delta F \) denotes flow adjustments (e.g., modified connectivity), and \( \Delta C \) captures content refinements (e.g., altered statements).

\subsubsection{Prompting the LLM with CFG Diff}  
The LLM is prompted with \( \Delta G \) to analyze the target code \( C_t \) and suggest optimizations. The prompt includes:

\textbf{Structural Insights:} Observations on execution paths and flow complexity based on \( \Delta G \), highlighting critical changes and their potential impact on program behavior.

\textbf{Optimization Recommendations:} Groundtruth patch, such as loop unrolling or branch simplification, annotated from common patterns in \( \Delta G \).

\subsubsection{Retrieval-Augmented Guidance}  
To enhance the LLM’s contextual understanding, a RAG pipeline retrieves examples \( E_r \) from a dataset \( D \), stored in a vector database. Each entry in \( D \) contains:

\begin{itemize}
    \item \textbf{Code Pairs:} (\( C_o, C_p \)), where \( C_o \) is the original code and \( C_p \) is its optimized counterpart, annotated with the corresponding \( \Delta G \) to capture critical transformations and their context.
    \item \textbf{Embeddings:} \( e_{cfg} \), derived from CFGs, representing the combined semantic and structural features of the code to facilitate similarity matching during retrieval.
    \item \textbf{Explanations:} Contextual rationales for the transformations, generated by the LLM based on \( \Delta G \) analysis, offering insights into optimization strategies.
\end{itemize}

Relevance is determined using cosine similarity: \( \text{Sim} = \cos(e_{cfg}^t, e_{cfg}^r) \), where \( e_{cfg}^t \) and \( e_{cfg}^r \) represent the embeddings of the target code and retrieved example, respectively. The vector database ensures efficient retrieval of examples that align with the structural and semantic context of the target code, guiding the LLM in producing optimized outputs.

\subsection{Combining Retrieved Insights for Code Optimization}

After computing the CFG differences \(\Delta G = (\Delta S, \Delta F, \Delta C)\) and compiling any associated optimization rationales, we incorporate these details into a structured prompt together with \emph{exactly one} retrieved example. Let us denote this prompt as:
\[
P = \bigl( \Delta G,\, R_{\text{opt}},\, E_r \bigr),
\]
where \(\Delta G\) captures structural differences, \(R_{\text{opt}}\) represents optimization rationales, and \(E_r\) is the single retrieved example. In practice, we rank potential examples by structural similarity and then pick the top candidate. Preliminary experiments retrieving two or three examples did not yield additional gains and often produced longer, less focused prompts.

\begin{itemize}
    \item \textbf{CFG Differences (\(\Delta G\)):} Descriptions of added or removed blocks (\(\Delta S\)), changes in control flow (\(\Delta F\)), and statement-level modifications (\(\Delta C\)). These clues highlight effective structural edits found in prior code pairs.
    \item \textbf{Optimization Rationales \((R_{\text{opt}})\):} Observations on why certain transformations (for instance, loop simplification or branch reduction) improve efficiency, helping the LLM replicate them in new contexts.
    \item \textbf{Retrieved Example \((E_r)\):} A training code pair with a closely related CFG, including both unoptimized and optimized versions. This shows how precise edits can enhance runtime performance.
\end{itemize}

By embedding CFG knowledge, the pipeline focuses on high-impact edits rather than simple text-level replacements. This CFG-based approach uncovers potential loop reorganizations, branch eliminations, or data-flow optimizations that can unlock meaningful speedups. Although retrieving only one example limits the variety of patterns, our empirical findings confirm that selecting a single, structurally close match effectively guides the LLM to context-relevant improvements.

\section{Experimental Design}

This section details the experimental setup, dataset, evaluation metrics, and results to validate the effectiveness of our proposed context-aware optimization pipeline. By integrating context-aware analysis with historical patch retrieval for LLMs, we evaluate its capability to optimize C++ code and achieve measurable improvements over baseline methods. 

\begin{table}[t]
    \centering
    \caption{Dataset Composition, sampled from the IBM dataset, with C++ as the primary language.}
    \label{tab:dataset_stats}
    \resizebox{0.7\columnwidth}{!}{
    \begin{tabular}{l|c}
        \toprule
        \textbf{Property} & \textbf{Value} \\
        \midrule
        Total code pairs & 1,200 \\
        Vector DB code pairs & 1,000 \\
        Testing code pairs & 200 \\
        \midrule
        Primary programming language & C++ \\
        Diversity of problems & High \\
        \bottomrule
    \end{tabular}}
\end{table}

\subsection{Experiment Preparation}  
Table~\ref{tab:dataset_stats} summarizes the dataset, drawn from the IBM Project CodeNet~\cite{puri2021codenet}, containing 1,200 C++ code pairs of source and optimized programs. The data is split into 80\% (1,000 pairs) for the vector database and 20\% (200 pairs) for testing. Experiments were conducted on a system equipped with an Intel Xeon Gold 6330N CPU featuring 28 cores, 56 threads, and 43 MB of cache, ensuring efficient processing and robust evaluation across various optimization scenarios.

\subsection{Baseline Methods}  
We evaluate \TheName{} against two baselines: \textit{Zero-Shot Generation}, which uses GPT-4o without context or retrieval, and \textit{Naive Generation}, which retrieves repair examples using source code embeddings. All methods share the same prompt structure to ensure fair performance comparison.

\subsection{Evaluation Metrics}

To comprehensively evaluate the pipeline, we adopt two categories of metrics: lexical similarity and execution time. Lexical metrics assess how closely the generated patches align with ground truth, while execution time metrics measure the practical impact of optimizations across different types.

\subsubsection{Lexical Similarity Metrics}
Lexical similarity metrics evaluate how closely generated patches match the ground truth, providing an initial assessment of patch fidelity:

\begin{itemize}
    \item \textbf{Line Overlap (LO):} Measures the percentage of matching lines between the generated code and the ground truth:
    \( LO = \frac{L_m}{L_t} \times 100 \),
    where \( L_m \) is the number of matching lines and \( L_t \) is the total number of lines in the ground truth.
    
    \item \textbf{Edit Distance Similarity (EDS):} Captures the closeness of generated code by comparing the minimum edit operations:
    \( EDS = 1 - \frac{E}{\max(L_g, L_t)} \),
    where \( E \) is the edit distance, \( L_g \) is the length of the generated code, and \( L_t \) is the length of the ground truth.
    
    \item \textbf{Token Overlap (TO):} Analyzes the similarity of token sequences by computing the percentage of matching tokens:
    \( TO = \frac{T_m}{T_t} \times 100 \),
    where \( T_m \) is the number of matching tokens and \( T_t \) is the total number of tokens in the ground truth.
\end{itemize}

\subsubsection{Execution Time Metrics}

\begin{table}[tb]
    \centering
    \caption{Optimization Types and Their Distribution in the Common Test Set. Each code sample may have multiple labels.}
    \label{tab:optimization_distribution}
    \resizebox{0.9\columnwidth}{!}{
    \begin{tabular}{l|l|c}
        \toprule
        \textbf{Optimization Type} & \textbf{Description} & \textbf{Count} \\
        \midrule
        Code Refactoring & Improves code structure & 303 \\
        Memory Optimization & Reduces memory use & 174 \\
        Performance Enhancement & Speeds up execution & 149 \\
        Algorithmic Simplification & Simplifies logic & 226 \\
        Loop Optimization & Boosts loop efficiency & 90 \\
        \midrule
        Total & - & 942 \\
        \bottomrule
    \end{tabular}}
\end{table}

Execution time provides a direct measure of the practical impact of generated patches, validating improvements in program performance beyond lexical similarity. By evaluating execution time across the fine-grained optimization types in Table~\ref{tab:optimization_distribution}, we gain deeper insights into how effectively the pipeline addresses specific challenges like execution speed, memory usage, and algorithmic complexity, demonstrating its versatility in diverse scenarios.

\subsection{Code Preprocessing and Model Selection}

To enable effective analysis, C++ code snippets are converted to CFGs using \textit{Clang's static analyzer}\footnote{\url{https://clang.llvm.org/docs/ClangStaticAnalyzer.html}}. Preprocessing standardizes headers, resolves dependencies, and removes unsupported attributes, creating structured inputs for analysis. The test set, refined to 116 executable programs from IBM Project CodeNet~\cite{puri2021codenet}, excludes non-executable or anomalous code. Execution testcases are generated by GPT-4o~\cite{hurst2024gpt} and embeddings are generated by pretrained CodeBERT~\cite{feng2020codebert}, ensuring consistency in retrieval and generation.

\section{Results and Analysis}

This section evaluates the performance of \TheName{} using lexical similarity and execution time metrics. These metrics offer complementary perspectives on how effectively the generated patches align with the ground truth and how much they improve runtime efficiency. By integrating CFG-based analysis, our approach addresses higher-level structural edits that pure text-based or naive retrieval methods can overlook.

\subsection{Lexical Similarity Analysis}

\begin{table}[t]
    \centering
    \caption{Lexical Comparison Metrics (\%). LO: Line Overlap, EDS: Edit Distance Similarity, TO: Token Overlap.}
    \label{tab:lexical_comparison}
    \resizebox{0.8\columnwidth}{!}{
    \begin{tabular}{l|c|c|c}
        \toprule
        \textbf{Generation Type} & \textbf{LO (\%)} & \textbf{EDS (\%)} & \textbf{TO (\%)} \\
        \midrule
        Zero-Shot Generation & 8.17 & 14.57 & 54.51 \\
        Naive Generation & 8.11 & 14.58 & 55.21 \\
        Context Generation & 8.53 & 16.54 & 59.91 \\
        \midrule
        \textbf{Improvement (\%)} & +4.41\% & +13.52\% & +9.91\% \\
        \bottomrule
    \end{tabular}}
\end{table}

As shown in Table~\ref{tab:lexical_comparison}, \TheName{} consistently outperforms baselines in LO, EDS, and TO. These improvements indicate that including CFG-focused prompts and relevant examples encourages the generation of patches more closely aligned with the structure and semantics of the optimized code. While higher lexical similarity does not guarantee logical correctness, it suggests that the approach captures patterns of effective edits that typically reflect deeper program analysis.

\subsection{Execution Time Analysis}

\begin{table}[t]
    \centering
    \caption{Average execution times and improvements for Zero-Shot, Naive, and Context Generation methods.}
    \label{tab:execution_times_and_improvement}
    \resizebox{0.8\columnwidth}{!}{
    \begin{tabular}{l|c|c}
        \toprule
        \textbf{Type} & \textbf{Avg Time (s)} & \textbf{Imp (\%)} \\
        \midrule
        Zero-Shot Generation & 0.4115 & - \\
        Naive Generation & 0.5238 & -27.3\% \\
        Context Generation & 0.3815 & +7.3\% \\
        \bottomrule
    \end{tabular}
    }
\end{table}

Table~\ref{tab:execution_times_and_improvement} shows that \TheName{} reduces execution time by 7.3 percent over zero-shot generation. Although modern compilers already eliminate many low-level inefficiencies, higher-level structural constraints often require human judgment. By integrating CFG insights, our method identifies small but critical edits, such as refining loop conditions or removing redundant code, that translate into tangible runtime benefits. This result highlights the importance of retrieving relevant examples that expose systematic changes, rather than relying solely on token-level patterns.

\subsection{Optimization Type Analysis}

\begin{figure}[t]
    \centering
    \includegraphics[width=\linewidth]{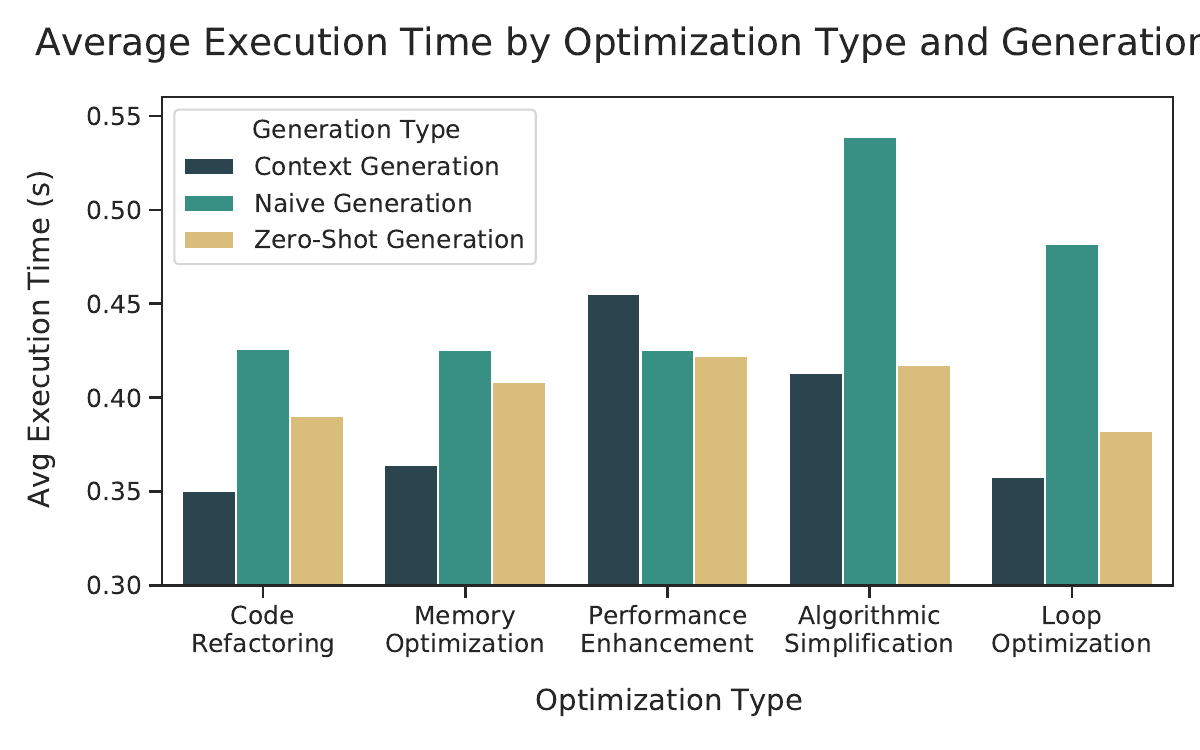}
    \caption{Average execution times for different optimization types across context-based, naive, and zero-shot generation methods. Context-based generation generally achieves lower execution times, except for a slight underperformance in performance enhancement.}
    \label{fig:execution_times_plot}
\end{figure}

Figure~\ref{fig:execution_times_plot} illustrates the average execution times for different optimization categories. \TheName{} demonstrates the lowest runtimes in tasks involving code refactoring, memory optimization, algorithmic simplification, and loop optimization. By leveraging CFG-based prompts, it better recognizes repetitive control flows and identifies specific opportunities for streamlining. A slight underperformance in the performance enhancement category suggests that certain specialized changes may require more domain-specific knowledge or an expanded set of retrieval examples. Even so, the overall adaptability across diverse challenges underscores the effectiveness of guiding large language models with structural context, enabling \TheName{} to produce optimizations that align more closely with real-world developer best practices. 

% \section{Discussion and Future Work}

% Future developments for \TheName{} aim to expand its role in software engineering. Its context-aware framework can support automated debugging by leveraging historical patterns for logical error resolution. CFG-driven analysis could be extended to program repair, proposing targeted patches for vulnerabilities and inconsistencies while addressing critical security issues. Additionally, \TheName{} holds potential for domain-specific performance tuning, balancing speed, resource usage, and maintainability in real-time applications.

\section{Discussion and Future Work}

Our CFG-guided framework demonstrates promising runtime gains, but adopting it in real-world systems introduces new considerations for maintainability and domain-specific constraints.

\textbf{Potential Maintainability Issues:}  
Although CFG-based edits can streamline control paths and speed execution, transformations such as loop unrolling or branch merging may affect readability. For example, expanded loops or fewer conditionals could complicate debugging by obscuring critical logic. Similarly, removing intermediate variables to optimize speed may reduce code clarity. Integrating maintainability metrics (such as cyclomatic complexity) and developer feedback could help balance short-term performance gains with long-term code upkeep.

\textbf{Future Directions:}  
Beyond optimization, the context-aware framework is well suited for advanced debugging, leveraging historical examples of logical error resolution. CFG-focused prompts also align with deeper program repair, including targeted security patches and domain-specific enhancements. Extending the retrieval corpus with specialized data and feedback loops would further broaden \TheName{}’s potential for safe, efficient, and comprehensible code transformations across diverse software engineering tasks.

% Ultimately, combining \TheName{} with advanced retrieval approaches, more sophisticated embeddings, and adaptive prompt engineering could broaden its utility to a wider range of programming tasks, from automated debugging and repair to large-scale runtime optimization.

\section{Threats to Validity}

% Our findings are based on a set of C++ programs drawn from IBM Project CodeNet, which may not capture the diversity of real-world codebases, programming languages, or hardware architectures. In addition, the retrieval process in our RAG pipeline could influence results depending on how many examples are fetched and how CFG knowledge is embedded. We only compared a few specific baselines (for instance, GPT-4o in zero-shot mode and a naive retrieval approach), so the improvements we observed might differ under alternative configurations, random retrieval elements, or more advanced prompting strategies.

% We primarily quantified performance gains through execution time improvements, complemented by lexical similarity metrics. Such measures, however, do not reflect maintainability, scalability, or project-specific constraints. Certain optimizations might also rely on hardware features or compiler behaviors not accounted for in our setup. Going forward, evaluating the generated code in broader, more realistic contexts, along with providing replication packages for deeper scrutiny, will be essential to strengthening the generalizability and practical relevance of our approach.

Our findings draw on C++ programs from IBM Project CodeNet, which may not represent the full diversity of real-world codebases, languages, or hardware. The retrieval process in our RAG pipeline (such as how many examples are fetched or how CFG knowledge is embedded) could also influence results. We only compared a few baselines (for instance, GPT-4o in zero-shot mode and a naive retrieval approach), so improvements might vary with different configurations or advanced prompting strategies.

We primarily measured performance gains through execution time, supplemented by lexical similarity metrics. However, these do not cover maintainability, scalability, or project-specific constraints. Some optimizations may also depend on hardware features or compiler behaviors outside our scope. Moving forward, evaluating generated code in broader contexts—and providing replication packages—will be vital for improving generalizability and relevance.

\section{Related Work}

We build on advances in human-centered AI, retrieval-augmented generation (RAG), in-context learning, and program analysis. By leveraging insights from developer attention studies and CFG-based analysis, our method addresses program-specific dependencies and structural nuances for more precise, context-aware code optimizations.

\textbf{Human-Centered AI for Neural Code Comprehension:}
Human-centric research highlights how developers and AI systems align when interpreting code. Eye-tracking and scanpath prediction reveal points where neural and human attention diverge, affecting explainability~\cite{li2024machines,bansal2023modeling,zhang2024eyetrans,karas2024tale}. Efforts to learn representations for source code and binaries further underscore how modeling human-like focus can enhance automated code understanding~\cite{zhang2022pre,zhang2022leveraging}. However, most of this work targets interpretability rather than deeper structural analysis (e.g., CFGs). Our approach extends these insights to optimize runtime, combining human-inspired reasoning with automated transformations.

\textbf{RAG and In-Context Learning for Code LLMs:}
Retrieval-Augmented Generation (RAG) enriches large language models with external knowledge, improving tasks like code generation, refactoring, and bug fixing~\cite{zhang2024rag,tan2024prompt,mozharovskii2024evaluating,li2024malmixer,bhattarai2024enhancing}. Frameworks such as ARKS~\cite{su2024arks} and CodeRAG-Bench~\cite{wang2024coderag} integrate structured and unstructured data but often overlook intricate syntactic dependencies~\cite{koziolek2024llm,fan2024survey,zhao2024retrieval}. Meanwhile, \emph{in-context learning} allows LLMs to adapt without fine-tuning by embedding labeled examples in prompts~\cite{zhang2023makes,wies2024learnability,ye2023compositional,li2023finding,kapu2024democraft}, yet struggles with complex control-flow reasoning~\cite{pawelczyk2023context,wies2024learnability}. By uniting RAG with in-context learning under a CFG-based framework, we provide explicit structural cues that yield more reliable, performance-centered code generation.

\textbf{Program Analysis for Optimization:}
Traditional program analysis uses static or dynamic techniques to address correctness, resource usage, and speed~\cite{fan2023static,chen2024evaluating,wen2024enchanting,erhabor2023measuring}. While ML-based methods can detect inefficiencies, many overlook context-dependent details in large-scale projects~\cite{zhang2024detecting,chew2023llm,nam2024using,bairi2024codeplan,rafi2024enhancing,wadhwa2023frustrated}. Minor loop or conditional edits can radically change execution behavior, underscoring the need for deeper structural insight. By embedding CFG analysis into an LLM workflow, we align semantic and syntactic contexts for more precise, performance-oriented optimizations that surpass traditional ML or heuristic-driven solutions.

\section{Conclusion}

In conclusion, this work introduces \TheName{}, an approach that combines in-context learning with CFG analysis to empower LLMs in generating optimized code effectively. Inspired by human cognitive processes, \TheName{} integrates historical examples and program-specific context, bridging the gap between retrieval-based methods and static analysis. Experimental results showcase a 7.3\% improvement in execution efficiency over GPT-4o, underscoring its ability to deliver consistent and meaningful optimizations. Future research will extend \TheName{}'s capabilities to applications like automated debugging, program repair, and domain-specific performance tuning, paving the way for a unified framework for intelligent and adaptable program optimization in diverse software engineering scenarios.

\section*{Data Availability Statement}

All data and code used in this study are available at 
\href{https://github.com/manishacharya60/rag-optimization}{\texttt{rag-optimization}}\footnote{\url{https://github.com/manishacharya60/rag-optimization}}. 
The repository includes:
\begin{itemize}
    \item \textbf{Preprocessed Datasets:} C++ code pairs, each with original and optimized variants, along with metadata on problem type and complexity. Also includes code embeddings used for retrieval.
    \item \textbf{Preprocessing and CFG Generation Scripts:} Python utilities to normalize input code, invoke Clang for control flow graph extraction, and parse CFG outputs into structured representations.
    \item \textbf{Implementation Modules and Instructions:} Source files for our retrieval-augmented approach, baseline methods, and step-by-step guidelines to replicate or extend the experiments.
\end{itemize}

Researchers are encouraged to explore these resources to replicate our results or adapt them for related investigations.

% \section*{Data Availability Statement}

% All data and code used in this study are publicly accessible at \texttt{rag-optimization}\href{https://github.com/manishacharya60/rag-optimization}{[Link]}\footnote{\url{https://github.com/manishacharya60/rag-optimization}}. 
% The repository contains:
% \begin{itemize}
%     \item Preprocessed datasets for training and evaluation.
%     \item Scripts for data preprocessing, CFG generation, and retrieval-augmented prompting.
%     \item Python modules (\texttt{code\_analysis.py}, \texttt{context\_aware.py}, etc.) implementing our optimization framework and baseline comparisons.
%     \item Configuration files and documentation to facilitate replication of experiments.
% \end{itemize}

% Researchers are encouraged to clone and explore the repository to replicate our findings or adapt these resources to their own projects. Questions or issues may be directed to the authors. 

\section*{Acknowledgments}

This work was supported in part by The SyBBURE Searle Undergraduate Research Program and by the National Science Foundation under Grant No.\ CCF-2211429. Any opinions, findings, and conclusions or recommendations expressed in this material are those of the authors and do not necessarily reflect the views of The SyBBURE Searle Undergraduate Research Program or the National Science Foundation.

\bibliographystyle{IEEEtran}
\bibliography{IEEEtran}

\end{document}